\documentclass[a4paper,10pt]{article}
\usepackage{a4wide}
\usepackage[utf8]{inputenc}
\usepackage[numbers]{natbib}
\usepackage{amsmath}
\usepackage{graphicx}
\usepackage{hyperref}
\usepackage[outdir=./]{epstopdf}

\title{Simulation of the COVID-19 pandemic on the social network of 
Slovenia: estimating the intrinsic forecast uncertainty}
\author{Žiga Zaplotnik, Aleksandar Gavrič, Luka Medic}

\begin{document}

\maketitle

\begin{abstract}
In the article a virus transmission model is constructed on a simplified social 
network. The social network consists of more than 2 million nodes, each 
representing an inhabitant of Slovenia. The nodes are organised and 
interconnected according to the real household and elderly-care center 
distribution, while their connections outside these clusters are semi-randomly 
distributed and fully-linked. The virus spread model is coupled to the disease 
progression model. The ensemble approach with the perturbed transmission and 
disease parameters is used to quantify the ensemble spread, a proxy for the 
forecast uncertainty.

The presented ongoing forecasts of COVID-19 epidemic in Slovenia are compared 
with the collected Slovenian data. Results show that infection is currently 
twice more likely to transmit within households/elderly care centers than 
outside them. We use an ensemble of simulations ($N=1000$) to inversely obtain 
posterior distributions of model parameters and to estimate the COVID-19 
forecast uncertainty. We found that in the uncontrolled epidemic, the intrinsic 
uncertainty mostly originates from the uncertainty of the virus biology, i.e. 
its reproductive number. In the controlled epidemic with low ratio of infected 
population, the randomness of the social network becomes the major source of 
forecast uncertainty, particularly for the short-range forecasts. 
Social-network-based models are thus essential for improving epidemics 
forecasting.
\end{abstract}

\section{Introduction}
The ongoing COVID-19 epidemic has revealed a major gap in our ability to 
forecast the evolution of the epidemic. There are several ways to simulate the 
epidemic dynamics. The most common approach is using compartmental models of 
susceptible (S), immune (I), recovered (R) population, i.e. SIR models 
\citep{Kermack1927,Hethcote2000}. These are described by a system of 
differential equations given some predefined parameters, such as probability of 
the disease transmission and the rate of recovery or mortality. Another 
variation of the SIR model is a SEIR model, which accounts also for the exposed 
(E) population, representing infected but not yet infectious subjects. The SEIR 
model is combined with activation functions to smoothly model social factors 
affecting virus spread and the disease progression. A major setback of the 
deterministic epidemic models is that they are only applicable for sufficiently 
large populations, where the assumption of homogeneous spread of the virus is 
valid \citep{Bartlett1957}. For coronaviruses, there is evidence that some 
infectious cases, the so called superspreaders, spread virus more than 
others \citep{Lloyd-Smith2005}. Their role is of the utmost importance in the 
initial uncontrolled phase of an epidemic and in its final controlled phase 
with low population of infected. Thus, deterministic SEIR models are also unable 
to properly describe the intrinsic uncertainty of the virus spread due to 
heterogeneous connectivity of the social network and due to heterogeneous 
disease progress of the infected population.

Consequently, we use node-based approach \citep{Eubank2010} to simulate 
the SARS-CoV-2 coronavirus spread over a simplified social network of more 
than 2 million nodes with a total of up to 20 million connections, representing 
the population of Slovenia and the connections of its inhabitants, with 
realistic distinction between household and outer connections. Despite being 
computationally more expensive, an advantage of network approach is that it 
allows direct simulation of intervention measures as well as planning of the 
future strategies of the the virus containment \citep[e.g.][]{Halloran2002} and 
lockdown-exit strategy. 
Here, we combine the network approach together with an ensemble-of-simulations 
approach which allows to estimate the uncertainty of the predictions which stems 
from the randomized network, initial number of infected, and the uncertainty of 
the coronavirus transmission parameters and disease progress parameters.

Section 2 describes the social network model, the virus transmission model and 
the coupled disease progression model. The probabilistic 
ensemble forecast of the COVID-19 pandemic for Slovenia and the contribution of 
different model components to the total forecast uncertainty is described in 
Section 3. Discussion, conlusions and further outlook are given in Section 4.

\section{Methodology}

\subsection{Social network model}
The social network model of the population of Slovenia distinguishes household 
connections and connections outside households. A total of $N=2045795$ nodes is 
used in the social network. The number of $k$-person households is given in 
Table~\ref{tab:household_num}. There are approximately 100 elderly care centers 
in Slovenia with a total of approximately 20000 residents. Each elderly care 
center is assumed to include 8 distinct groups of 25 people. Average 
household/care group consists of 2.5 people in Slovenia so the average number 
of contacts per person within household is 1.5. 

\begin{table}
\begin{center}
\begin{tabular}{ |l|l| }
 \hline
 $k$ persons in household & number of $k$-person households \\
  \hline
 1 & 269898 \\ 
 2 & 209573 \\ 
 3 & 152959 \\
 4 & 122195\\
 5 & 43327\\
 6 & 17398\\
 7 & 6073\\
 8 & 3195\\
 \hline
 25 & 100 care centers with 8 groups each \\
 \hline
\end{tabular}
\caption{\label{tab:household_num}Households distribution in Slovenia, based on 
the data of Statistical Office of Republic of Slovenia \citep{Dolenc2018}.}
\end{center}
\end{table}

\begin{figure}
 \centering
\includegraphics[width=0.7\textwidth]
{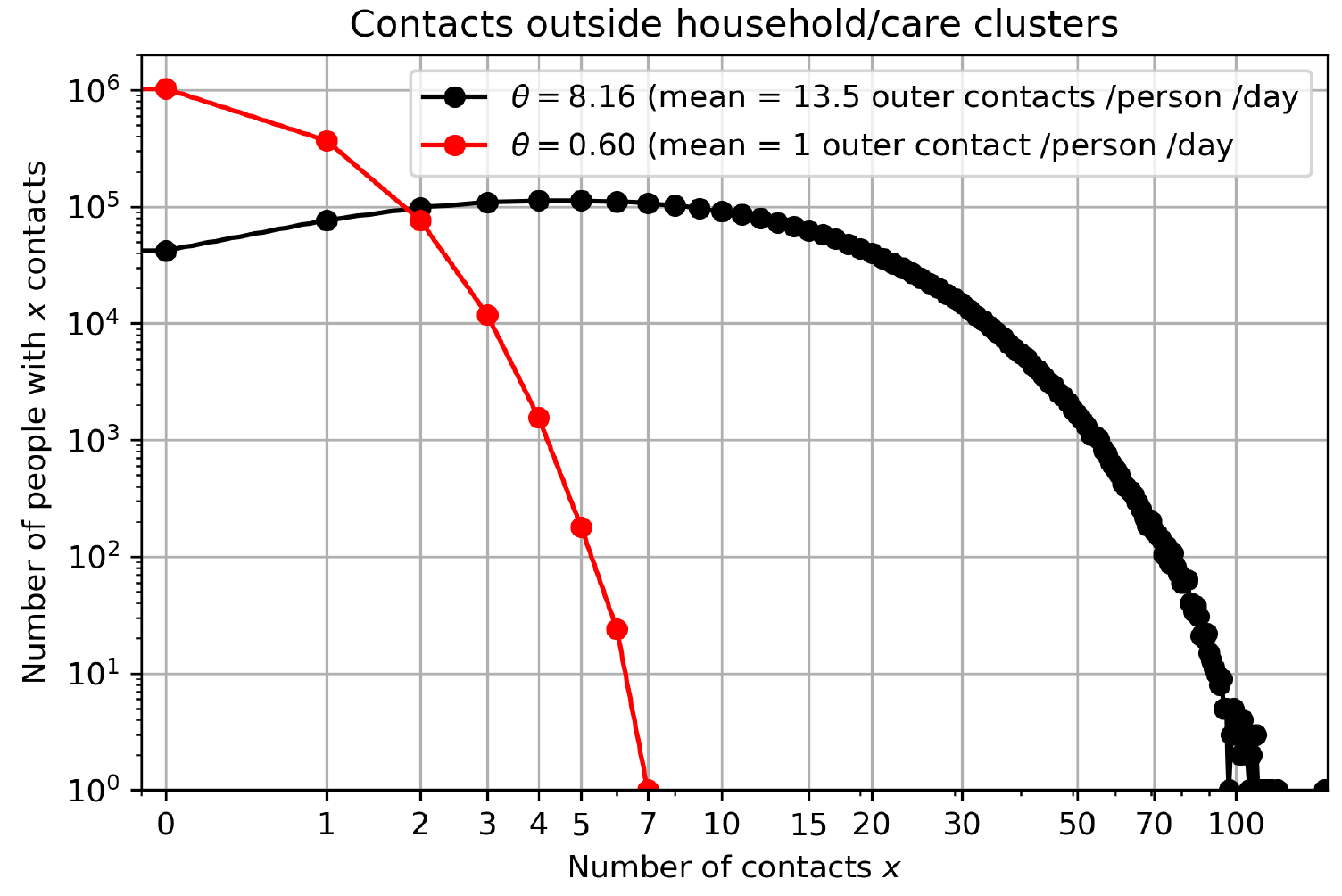}
\caption{Number distribution $N(x) = p(x) N$ by the number of contacts in the 
social network model. The black graph shows the assumed distribution of people 
with a given number of outer contacts in a normal, non-epidemic, phase, while 
the red graph presents reduced number of outer contacts in the case of a social 
distancing measures.}
\label{fig:contacts_distribution}
\end{figure}

In normal conditions, contact number distribution follows power law with fat 
tails \citep{Adamic2001}, which are associated with superspreader events, e.g. 
large public gatherings such as sport and cultural events. 
However, since all public events are canceled in the event of the COVID-19 
epidemic, these fat tails are cut off \citep{Norman2020} and the topology of the 
social network changes substantially. In conditions without large public 
gatherings, it is reasonable to assume that certain people still have much 
larger number of contacts than others. The studies of social mixing, e.g. 
POLYMOD study of social interactions within 8 European countries, typically 
report negative binomial distribution of the number of contacts 
\citep{Leung2017,Mossong2008}. We assumed mean number of 
contacts outside households to be 13.5 with standard deviation of 10.5. Instead 
of negative binomial distribution, we rather use smooth gamma distribution, 
which resembles the shape of the binomial distribution but has some 
useful mathematical properties, which will be exploited in the continuation. 
Thus, we model the connectivity, 
i.e. the number of contacts per person, using the gamma probability 
distribution, which is essentially an exponential distribution
\begin{equation}\label{eq:gamma_distribution}
 p(x; k,\theta) = \frac{1}{\Gamma(k) \theta^k} x^{k-1} e^{-\frac{x}{\theta}} .
\end{equation}
In this study, we use $k=1.65$ and $\theta=4.08$ for the initial setup, which 
gives an average number of 13.5 contacts per person per day 
(Fig.~\ref{fig:contacts_distribution}). Together with 1.5 family contacts per 
person per day, the total number of contacts per person per day is 15. Here, we 
assume that the average number of contacts is the same for each age group, 
despite studies showing that elderly have reduced number of contacts 
\citep{DelValle2007}. The average contact number per person per day varies for 
different countries \citep{Mossong2008}, however 15 contacts per day is 
a reasonable guess for Slovenia. We also assume quasi-static social network, 
i.e. only 20\% of contacts are new every day, and the remaining 80\% are static. 
This choice is a first guess, justified by the fact that only around 20\% of 
all daily contacts last less than 15 minutes \citep{Mossong2008}. These can be 
regarded as random sporadic contacts. Self-distancing measures to mitigate 
COVID-19 can be imposed by decreasing parameter $\theta$, which also decreases 
the average number of outer contacts (Fig.~\ref{fig:contacts_distribution}).

\begin{figure}
 \centering
\includegraphics[width=1.\textwidth]
{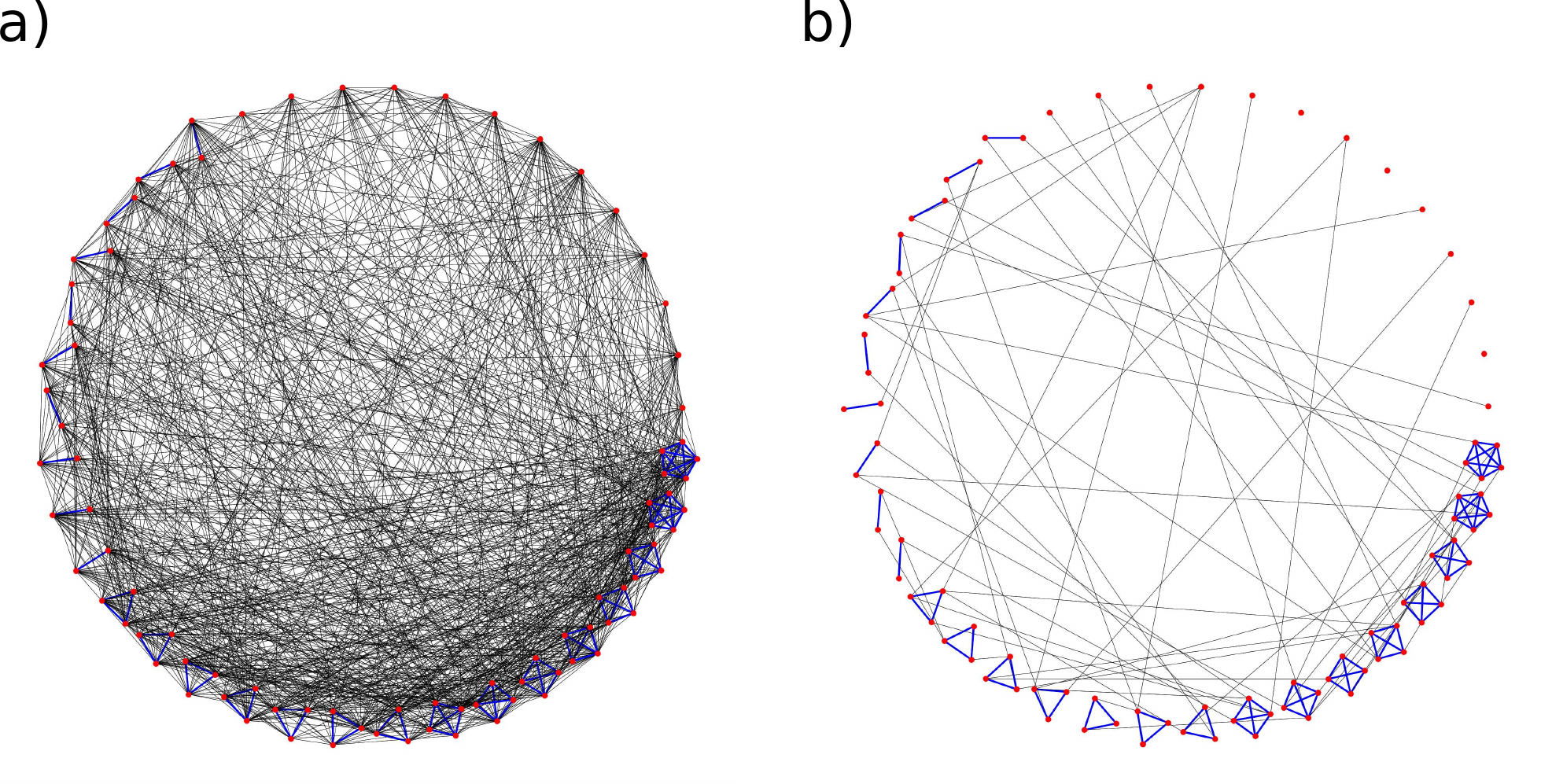}
\caption{Connectivity of the social network of $N=88$ nodes for a) densely 
connected graph, where each node has on average 15 contacts per day (1.5 
family and 13.5 outer contacts, $\theta=22.5$) and b) sparsely connected 
graph, where each node has on average 2.5 contacts per day (1.5 family and 1 
outer contact). Red dots are nodes, blue lines represents household connections 
and black lines outer connections. The graph represents a minimized version of 
the social network used in the virus spread simulation.}
\label{fig:graph}
\end{figure}

Fig.~\ref{fig:graph} shows an example of the connectivity change of a minimised 
network with 88 nodes clustered on a circle with the real household 
distribution taken into account.

Technically, we connect the graph in the following way:
\begin{enumerate}
 \item number of outer contacts for each node is randomly drawn from Gamma 
distribution (\ref{eq:gamma_distribution}). If node $i$ has $x_i=0.33$ contacts 
per day, it means that it will have 0 contacts 2/3 of the time and 1 
contact 1/3 of the time of the simulation;
 \item for each node $i$, we randomly assign the connections to $x_i$ other 
nodes, where $x_i$ is the number of contacts of node $i$. However, not every 
node has the same probability of being picked as a neighbour. Node $j$, which 
has $x_j$ contacts, is picked as a neighbour with probability $ x_j N(x_j)/T$, 
where $N(x_j)$ is the number of nodes with $x_j$ contacts and $T$ is the total 
number of contacts in the network ($T$ is twofold the number of connections). 
Sampling over Gamma distribution (\ref{eq:gamma_distribution}) gives us a 
distribution of $N(x)=p(x) N$. When picking the neighbours, we actually sample 
the same Gamma distribution times $x$, i.e.
\begin{equation}
 p_n(x)=p(x; k,\theta) x = \frac{1}{\Gamma(k) \theta^k} x^k 
e^{-\frac{x}{\theta}} \propto p(x; k+1,\theta) .
\end{equation}
 \item The shape of the social network is changing (80\% of connections 
static, 20\% changing) at every timestep of the simulation to account for 
random sporadic contacts. (a) The number of contacts of node $i$ is fixed 
(randomly jumps between $\lfloor x_i \rfloor$ and $\lceil 
x_i\rceil$ based on the value of $x_i$). For example, if a node has 0.33 
contacts per day, 1 contact is picked with probability 1/3 and 0 contacts
with probability 2/3. (b) The social network is partially rewired at every 
timestep to account for superspreaders mobility.
\end{enumerate}

An important advantage of our network approach is that the nodes are not 
connected randomly through ``half-links'' (directed connections linking egos 
to their contacts, the alters) \citep{Danon2011,Read2008}, such as in the vast 
majority of modelling studies, where the WAIFW 
(who-acquires-infection-from-whom) matrices were constructed based on the 
egocentric data. Instead, nodes are fully-linked.

\subsubsection{Compartments}

Similarly as in the deterministic SEIR model, we divided the population into 
compartments, where we used the following compartmental division: susceptible, 
infected, infectious (exposed), hospitalised with severe illness, hospitalised 
and critically ill (ICU treatment), hospitalised and fatal, and recovered. The 
latter are assumed to be immune for the whole simulation period. In the 
network model, a susceptible node becomes exposed (infected) with a certain 
probability (called an attack rate) when in contact with an infectious node. 
After a certain period of time, the infected node progresses into infectious 
state. In the accordance with the chosen compartmental division we monitor each 
node by adapting its clinical state at every timestep. Since all the periods 
are stohastic variables, e.g. infectious and recovery periods, those periods 
widely vary among the nodes (and the variations would be even greater if we 
would account for the demographic properties, e.g. age and sex).

\subsection{Virus transmission model}\label{subsection:virus_transmission}

\subsubsection{Reproduction number $R_0$}
\begin{figure}[h!]
 \centering
\includegraphics[width=1.0\textwidth]{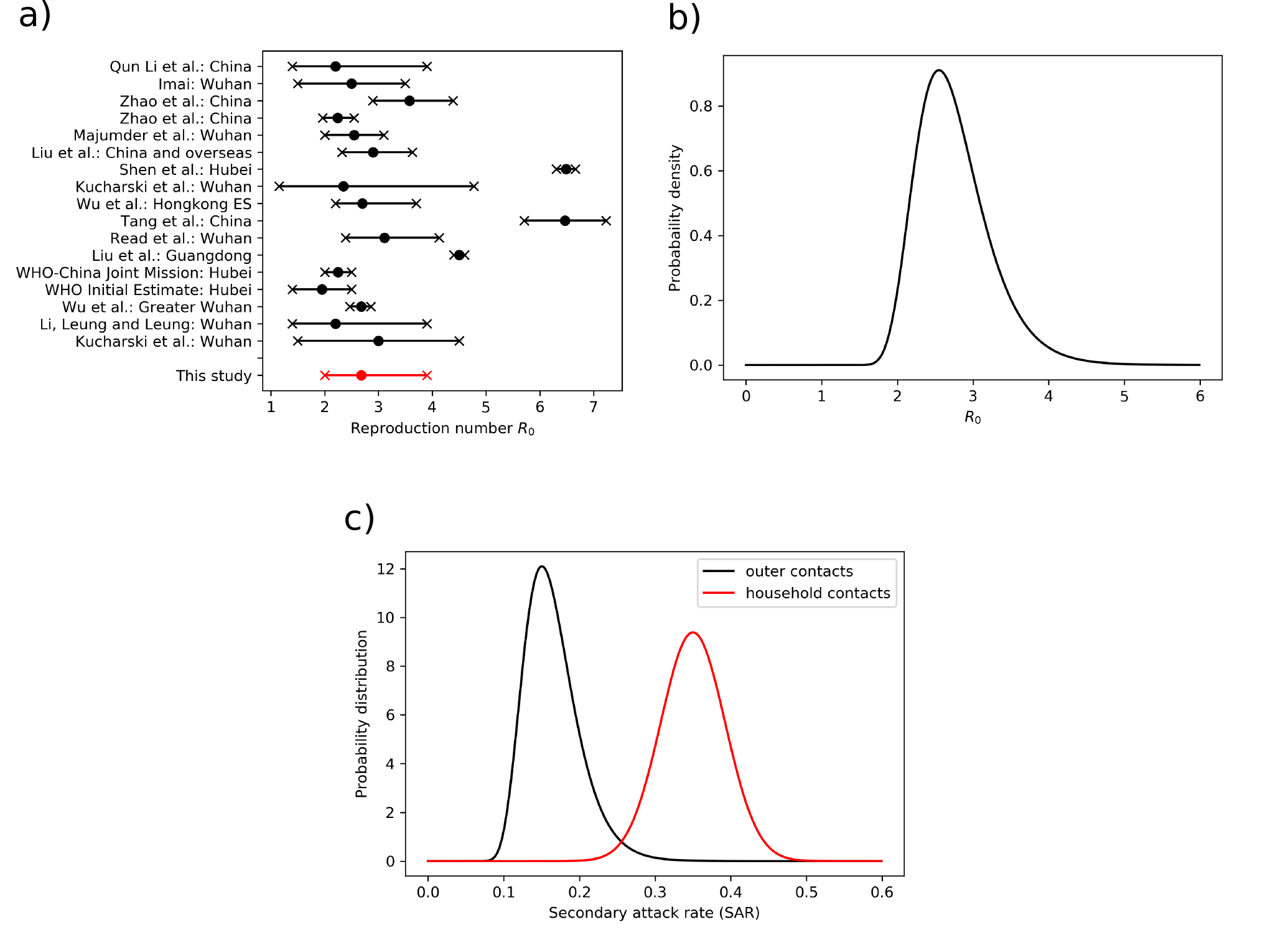}
\caption{a) Basic reproduction number $R_0$ (median and 95\% confidence 
interval), reported in a number of studies for different 
locations 
\citep{Liu2020a,Read2020,Kucharski2020a,Wu2020a,Li2020,Shen2020,
Majumder2020,Zhao2020,Imai2020,Tang2020,WHO2020} and references therein. b) 
Log-normal probability density function of the basic 
reproduction number, used for ensemble simulations. c) Probability distribution 
of secondary attack rates for household contacts and outer contacts.}
\label{fig:r0}
\end{figure}

A basic reproduction number, $R_0$, provides information on the average 
speed of virus transmission in an uncontrolled phase of the epidemic. Different 
methodologies produced different results, however the majority of reported $R_0$ 
for SARS-CoV-2 is within 2 and 3. Here, we use median reported $R_0$ from a 
number of studies, as well as its median confidence intervals, i.e. $R_0=2.68$, 
with 95\% confidence interval $[2,3.9]$. This approach is not the optimal one, 
since we are trading accuracy for precision. The published $R_0$ values as well 
as our deduced $R_0$ distribution is shown in Fig.~\ref{fig:r0}a,b. The 
optimal log-normal distribution should thus match the following conditions:
$\mathrm{CDF}(R_0^L;\mu,\sigma,\Delta x)=0.025$, 
$\mathrm{CDF}(R_0^U;\mu,\sigma,\Delta x)=0.975$, and $\mathrm{median (CDF) = 
R_0}$,
where $R_0^L$ and $R_0^H$ are lower and upper boundaries of $R_0$, CDF stands 
for log-normal cumulative distribution function
and its median is $\exp(\mu)$. Then, we define a quadratic cost function, 
which includes all the above criteria, and by minimizing it, we obtain the 
optimal parameters for log-normal distribution: $\Delta x=0.36, \sigma=1.14, 
\exp{\mu}=1.54$.

\subsubsection{Attack rate}
In general, the basic reproduction number, $R_0$ can be decomposed into the 
secondary attack rate times the number of contacts. The secondary attack rate 
($SAR$) is defined as the probability that an infection occurs among susceptible 
people within a specific group (i.e. household contacts or other contacts 
outside households). The measure can provide an indication of how social 
interactions relate to the transmission risk. We can further decompose the $R_0$ 
into the household risk of infection and outer risk of infection (following 
\citep{Liu2020})
\begin{equation}\label{eq:r_0}
 R_0 = SAR_h N_h + SAR_c N_c ,
\end{equation}
where $SAR_h$ and $SAR_c$ are secondary attack rates within household and 
outside household (outer contacts), respectively. $N_h$ and $N_c$ are the 
numbers of household contacs and outer contacts. Here, one must notice that the 
above estimation of $SAR_h$ assumed homogeneous mixing, while the network model 
is heterogeneous with $R_0 = SAR_h (\left<N_h\right> + 
\textrm{Var}(N_h)/\left<N_h\right>) + SAR_c (\left<N_c\right> + 
\textrm{Var}(N_c)/\left<N_c\right>)$. To be consistent with \citet{Liu2020}, we 
stick 
with formulation (\ref{eq:r_0}).

The study of \citet{Liu2020} 
suggest $SAR_h$ value of 35\% (95\% CI 27-44\%) for SARS-CoV-2. $SAR_h$ is 
almost normally distributed with mean 35\% and $2\sigma\approx 8.5$. The 
distribution of $R_0$ is given in the previous paragraph. It holds: 
$SAR_c=(R_0-SAR_h N_h)/N_c$. This gives a transmission efficiency of 
$SAR_c=$ 16\% (95\% CI 10.8-25.1\%). Fig.~\ref{fig:r0}c shows probability 
distributions of secondary attack rates as used in the ensemble of simulations.

If the social infectious period is $T_{inf}\approx$ 5 days (check 
subsection~\ref{subsubsection:infectious}), we can assume that the daily risk of 
getting infected from a certain household member is $SAR_{h,daily}$ where 
$1-(1-SAR_{h,daily})^{T_{inf}}=SAR_h$ and 
\begin{equation}
 SAR_{h,daily} = 1-\exp{\left( \frac{\ln{(1-SAR_h)}}{T_{inf}} \right)} 
\end{equation}
being equal 8.3\% (95\% CI 6.1-10.9\%). Similarly, we compute $SAR_{c,daily}=$ 
3.4\% (95\% CI 2.3-5.6\%).

Some studies \citep[e.g.][]{Burke2020} have concentrated only on the 
symptomatic secondary attack rates and have shown relatively smaller numbers: 
0.45\% (CI=0.12\%-1.6\%) among all close contacts and 10.5\% (CI=0.12\%-1.6\%) 
among household members. However, these numbers cannot reproduce the reported 
$R_0$ between 2 and 3.9 with any realistic number of contacts. Another study 
shows similar attack rates to what we use here \citep{Bi2020}.

The attack rate affects the virus transmission as follows. At each timestep of 
the simulation (every 1 day), the susceptible contacts of each infectious 
individual are randomly infected with probability $SAR_{h,daily}$ or 
$SAR_{c,daily}$, depending whether the contact occurs within household or 
outside it.

\subsection{Disease progression model}\label{subsection:disease}

\begin{figure}
 \centering
\includegraphics[width=1.0\textwidth]
{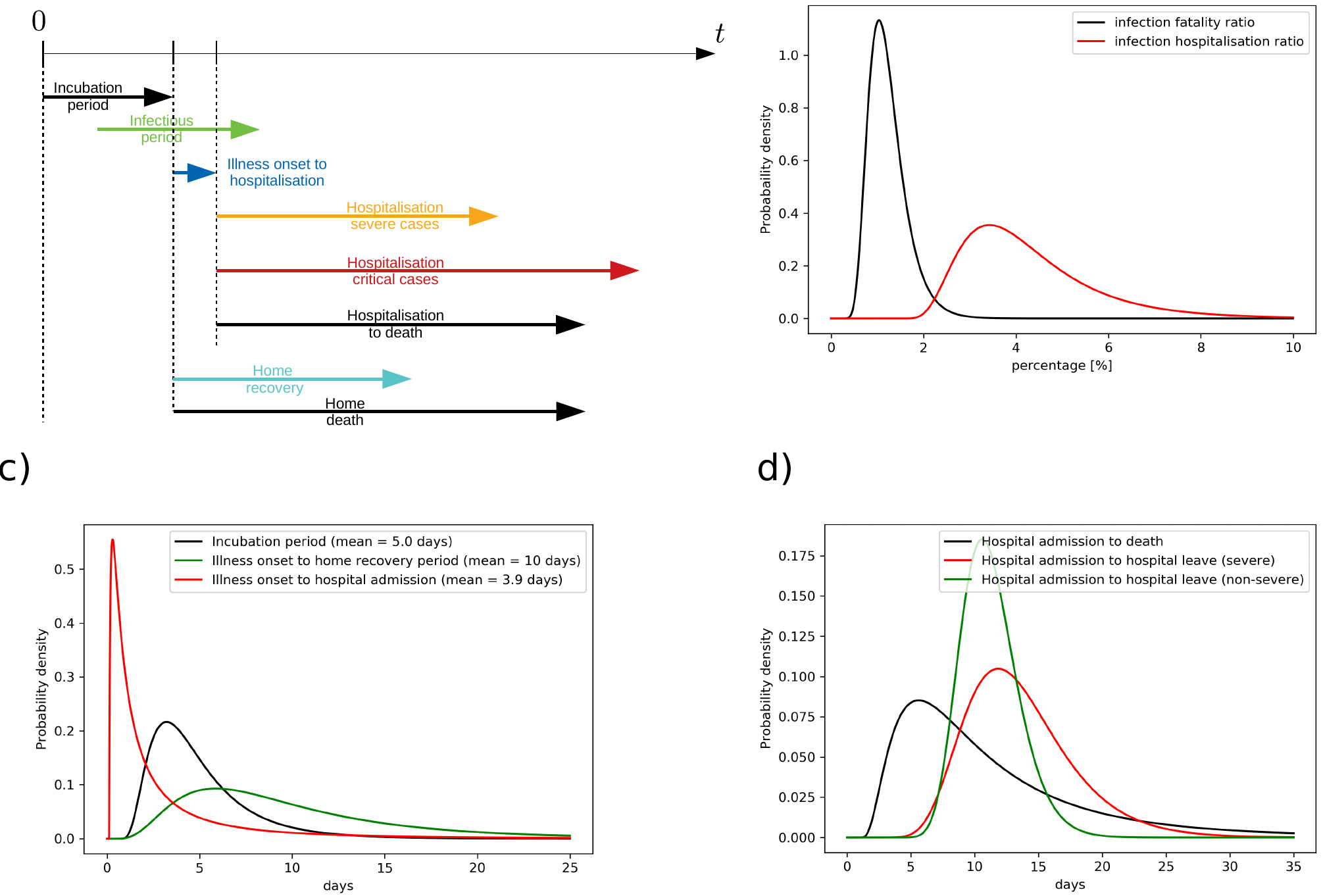}
\caption{a) A simplified sketch of illness evolution. b) Infection fatality 
ratio 
distribution and infection hospitalisation ratio distribution for ensemble 
simulations. Computed based on data from \citet{Verity2020}. c) Incubation 
period and illness onset to hospitalisation distribution for COVID-19 patients 
\citep{Linton2020}. d) Mean distribution of hospital admission to death, 
hospital admission to hospital leave for severe and for non-severe illness 
\citep{Linton2020,Guan2020a}.}
\label{fig:composite_disease}
\end{figure}

A simplified sketch of the disease progression model is shown in 
Fig.~\ref{fig:composite_disease}a. When a certain individual (node) gets 
infected, incubation period starts and several days will pass until the symptom 
onset. The majority of the infected people recovers at home, some die at home, 
while certain individuals are admitted to hospital in the following days. 
Several outcomes are possible: recovery after normal hospitalisation, recovery 
after intensive care unit hospitalisation and death. Note that for every node, 
the illness evolves differently based on the probability distribution described 
in the following subsections.  

\subsubsection{Case fatality ratio}
The baseline case fatality ratio (CFR), i.e. the fatality ratio among all 
positively tested, is assumed 1.38\% (CI 1.23-1.53\%) 
\citep{Verity2020,Russell2020}, similar to the estimate of \citet{Shim2020} for 
South Korea. Dividing deaths-to-date by cases-to-date leads to a biased estimate 
of CFR, called naive CFR (nCFR) as the delays from confirmation of a case to 
death is not accounted for, as well as due to under-reporting of cases and even 
deaths. The reported numbers agree with recently published study for symptomatic 
case fatality ratio in China \citep{Wu2020}.

\subsubsection{Infection fatality, intensive care and hospitalisation 
ratios}
Infection fatality ratio (IFR) estimates are based on the study from 
\citet{Verity2020}, which reported IFR of 0.66\% with 95\% confidence interval 
0.4\% to 1.3\%. These estimates are consistent with IFR on Princess Diamond 
Cruise ship, when demographic differences are accounted for 
\citep{Russell2020a}. In Imperial College report on COVID-19 
\citep{Ferguson2020}, 
these numbers have been also adjusted for the non-uniform attack rate and UK 
demography. The authors obtained age-stratified IFR estimates by adjusting their 
CFR estimates using COVID-19 prevalence data for expatriates evacuated from 
Wuhan. This approach involves very large uncertainties. Furthermore, 
\citet{Verity2020} collected data from patients who were 
hospitalised in Hubei, mainland China, where median age is 37.4 years while 
median age in Slovenia is 44.5 years. Study reported a strong age gradient in 
risk of death. We have applied those age-stratified 
estimates to the Slovenian population. Performing an age-stratified weighted 
average, we compute the total IFR of 1.16\% (CI 0.63-2.22\%). Similar total IFR 
was reported by a comprehensive study for Italy \citep{Rinaldi2020} (1.29\%, 
95\% CI 0.89-2.01\%). On the other hand \citet{Modi2020} have recently 
estimated somewhat lower IFR (0.95\%, 95\% CI 0.47-1.70\%) with the lower bound 
of 0.65\% for Lombardia, consistent with 0.58\% lower bound for Bergamo 
province. \citet{Villa2020} reported a bit higher IFR of 1.6\% (95\% CI 
1.1-2.1\%). A more comprehensive meta-analysis of IFR estimates was done by 
\citet{Meyerowitz-Katz2020}.

Analogously, we compute the average hospitalisation rate of 6.37\% (95\% CI 
3.8-13\%) based on \citet{Verity2020}. Slightly lower age-dependent 
hospitalisation rates were estimated for COVID-19 patients in USA by 
\citet{Garg2020}, which  adjusted for demography of Slovenia (but not 
accounting for non-uniform attack rate) gives hospitalisation rate of 3.97\%. 
The latter result better coincides with the observed number of 
hospitalisations in Slovenia. No interval estimate is given, thus we use the 
same relative error as given by \citet{Verity2020}. The final hospitalisation 
rate is thus 3.97\% (95\% CI 2.37-8.10\%). We assume that roughly 1/4-1/3
of all hospitalised cases are admitted to ICU \citep{Bialek2020}, despite 
some studies showing smaller proportions \citep{Chow2020}. We assume that one 
half of cases admitted to intensive care unit (ICU) are fatal 
\citep{ICNARC2020}.

Taking into account infection fatality ratio, hospitalisation ratio and ICU 
admission ratio, it follows that roughly one half of all deaths occur at 
home/elderly care center/palliative care center, which agrees with the present 
data for Slovenia \citep{sledilnik2020}. Note that for simplicity, we have 
assumed uniform attack rate across all ages, despite studies showing that 
working population is most likely to get infected 
\citep{Mizumoto2020a,Davies2020}. Using the minimization procedures, we obtain 
parameters of log-normal distribution which best fits both values and their 95\% 
confidence interval 
(Fig.~\ref{fig:composite_disease}b).



\subsubsection{Incubation period - infection to illness onset}
Mean incubation period is taken to be 5.0 days (95\% CI 4.2-6.0 days), while 
the 95th percentile of the distribution was 10.6 days (95\% CI 8.5-14.1 days) 
and 99th percentile 15.4 days (99\% CI 11.7-22.5 days) \citep{Linton2020}. 
Similar numbers were reported in earlier studies with less patients included 
\citep{Li2020,Liu2020b,Lauer2020,Kraemer2020}. Log-normal distribution is used 
among for incubation period among nodes. However, the parameters of the 
lognormal distribution remain fixed due to numerical instability of their 
computation. Thus, all ensemble members have the same log-normal distribution of 
incubation period. Incubation period distribution and other outcome parameters 
are shown in Fig.~\ref{fig:composite_disease}c.

\subsubsection{Infectious period}\label{subsubsection:infectious}
The infectious period is not yet well defined. A small study from German cohort 
of only 9 patients with mild clinical courses \citep{Woelfel2020} showed that 
viral shedding was high during the first week of symptoms and peaking at day 4. 
Another study from Singapore reported seven clusters in which virus was 
transmitted from a COVID-19 patient before experiencing symptoms. According to 
the authors pre-symptomatic transmission occurred 1-3 days before symptoms 
onset 
\citep{Wei2020}. We have therefore estimated latent (non-infectious) period of 
3 days and infectious period to start 2 days before the 
completion of incubation period (average incubation period is estimated at 5 
days). Thus, we assume 2 days of pre-symptomatic transmission. Slightly larger 
numbers (2.55 days for Singapore and 2.89 days for Tianjin, China) were 
reported by \citet{Tindale2020}. 

The infectious period likely 
ends around 5 days from symptoms onset, so the total period of infectiousness 
lasts $T_{bio inf}\approx$ 7 days. Note however that none of the interval 
boundaries are known exactly. Determining its final boundary is especially 
challenging, as it depends on the social factors as well, e.g. whether the 
infected cases are able to self-isolate from surroundings and how strictly they 
follow the self-isolation order. Here, we assume strict (100\%) self-isolation 
and use a \textit{social} infectious period of $T_{soc inf}$ = 5 days. It 
starts 2.5 days (95\% CI 1.5-3.5 days) after infection and end 2.5 days after 
incubation (95\% CI 1.5-3.5 days) as the case ascertainment typically occurs 2 
days after symptoms onset \citep{Moss2020}. The infectious period fall in line 
with study of \citet{Bi2020}, Fig. S2. It also falls in line with the 
reported proportion of pre-symptomatic transmission (representing half of 
infectious period) being 48\% for Singapore, 62\% for Tianjin, China 
\citep{Ganyani2020} and 44\% for 77 infector-infectee pairs in Gaungzhou, China 
\citep{He2020}.

\subsection{Illness onset to hospitalisation or home recovery}

From the illness onset on, there are two possible recovery pathways: home 
recovery or hospitalisation (Fig.~\ref{fig:composite_disease}c). Home 
recovery period for mild cases has not been documented officially but is 
reported to be within one and two weeks. Since it does not affect the 
hospitalisation statistics, we here assume it to be log-normally distributed 
with mean period of 10 days. 

Based on the clinical study of \citet{Linton2020}, mean illness onset to 
hospital admission period is 3.9 days (95\% CI 2.9-5.3 days), with median of 
1.5 days (95\% CI 1.2-1.9 days), 5\% percentile at 0.2 days (95\% CI 0.1-0.3 
days) and 95\% percentile at 14 days (95\% CI 10.3-20.1 days). Only the 
distribution of data for living patients is accounted for, since we now 
understand the severity of the illness. In China, fatal cases were admitted to 
hospital on average two days later \citep{Linton2020}.

\subsection{Hospital admission to recovery or death}

Hospital admission to death median (mean) length is assumed 6.7 (8.6) days long 
(Fig.~\ref{fig:composite_disease}d). Only slightly longer periods were 
reported by \citet{Mizumoto2020b} with mean length of 10.1 days. Hospital 
admission to recovery is on average longer than hospital admission to death. The 
median hospitalisation length is 11 days (95\% CI 10-13) for non-severe cases 
and 13 days for severe (95\% CI 11-17) \citep{Linton2020}. Both are 
log-normally distributed. For ensemble computations, their medians are further 
log-normally distributed according to their respective confidence intervals. 
Similar numbers were reported by \citet{Zhou2020} with 11 day (95\% CI 7-14) 
mean hospital length of stay and 8 day (95\% CI 4-12) mean ICU length of stay.

Fatality ratio of severe cases in need of intensive care is reported to be 
around 50\%. We assume fatality ratio of severe cases without intensive care to 
be normally distributed with mean of 90\% (95\% CI 85-95\%). Fatality ratio of 
severely ill without oxygen is assumed 10\% (95\% CI 5-15\%).

\subsection{Initial condition}
The initial condition for the simulation is defined for March 12, 2020. To that 
day, there were 131 symptomatic cases who tested positive in Slovenia, 8 days 
after first positive case, which implies an anomalously low doubling time of 
$\tau=$ 1.23 days. This number is case specific as there was winter holiday in 
Slovenia at the end of February and beginning of March. Thus, lots of cases were 
imported from Northern Italy (including Lombardy). Other studies typically 
suggest a doubling time of around 5 days (95\% CI 4.3 - 6.2) in the initial 
uncontrolled stage of the epidemic 
\citep{Ferretti2020a}. However, \citet{Abbott2020} report smaller values of 
around 3.5 days in most of Western Europe. Thus, our choice is doubling time of 
$T_{double}=$ 3.5 days (95\% CI 2.5-4.5 days) for the period before March 12.

Different numbers of actually infected people were suggested in the media 
reports, ranging from 5 to 20 times the number of reported positive cases. 
Given the average incubation period of 5 days + (2 days for case ascertainment) 
and doubling period of 3.5 days, factor $2^{\frac{T_{inc}+2}{T_{double}}}=4$ 
applies. Furthermore, the proportion of asymptomatic cases is around 18\% based 
on the data from Diamond Princess Cruise Ship \citep{Mizumoto2020} (mostly 
older people) and around 33\% based on the more recent study 
\citep{Nishiura2020}. Population screening tests from Iceland reported 41.6\% 
of all who tested positive, did not experience any COVID-19 symptoms 
\citep{Gudbjartsson2020}. Similar asymptomatic ratio (43.2\% (95\% CI 
32.2-54.7\%)) was reported also from a screening study conducted for the 
Italian town Vo \citep{Lavezzo2020}. Another study on the homeless population in 
Boston reported even larger proportion of asymptomatic cases 
\citep{Baggett2020}. The model-driven study of \citet{Emery2020} found that 
74\% (95\% CI 70-78\%) of SARS-CoV-2 infections proceeded asymptomatically, 
raising also doubts about the IFR. In this study, we opt for 40\%, normally 
distributed with standard deviation of 10\%. Furthermore, we double the value to 
account for the initial under-reporting of symptomatic cases, estimated by 
\citet{Kucharski2020a}. All together, this results in almost 1800 infected 
people in Slovenia by March 12. 

\begin{figure}
 \centering
\includegraphics[width=0.48\textwidth]
{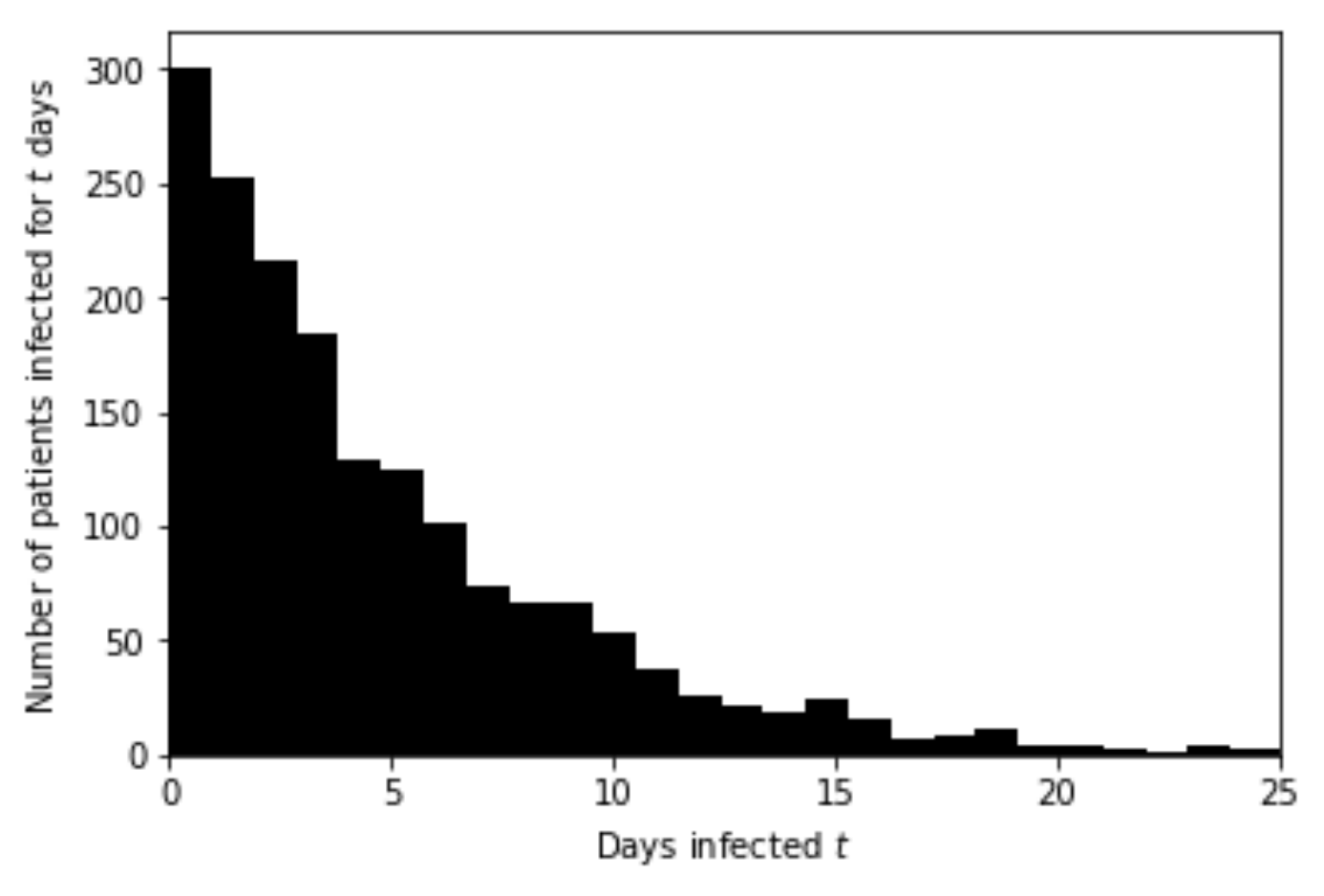}
\caption{Distribution of 1780 infected people on March 12, 2020, in Slovenia, 
by the duration of their infection.}
\label{fig:ic}
\end{figure}

Based on the exponential growth in the initial stage of the epidemic and known 
incubation period, we randomly generate the infection length of the 
patients with exponential distribution with shape factor of 
$T_{double}/\log{2}$, so that 131 develop symptoms and are ascertained by March 
12. Initial distribution of 1780 infected people by the time-length of their 
infection is shown in Fig.~\ref{fig:ic}. Note that in reality, due to many 
imported cases, the actual infection-time distribution may be slightly 
different.

\subsection{Ensemble of simulations}
Ensemble of simulations allows to estimate the uncertainty of the epidemic 
forecasts and to infer confidence in those predictions. There are two levels of 
perturbations in the ensemble: 1) at the 
start of each simulation, we perturb parameters, which govern the probability 
distributions of all model parameters, and 2) each node in the 
network has its own transmission 
probability based on its number of contacts and its own disease progression 
drawn from the associated probability distributions.

The uncertainty is associated with the impact of the intervention-measures on 
the social network connectivity and the uncertainty attributed to the intrinsic 
(internal, natural) model uncertainty. The latter can be further divided into:
\begin{enumerate}
 \item social network uncertainty associated with randomized connections,
 \item initial condition uncertainty as random nodes are infected,
 \item virus transmission dynamics uncertainty which stems from uncertainty 
of the parameters, described in Subsection~\ref{subsection:virus_transmission},
 \item disease progression model uncertainty due to uncertainty of the 
parameters, described in Subsection~\ref{subsection:disease}.
\end{enumerate}

In Slovenia, the intervention measures were imposed at several time 
instances between March 13 and March 30 \citep{Manevski2020}. Their impact was 
assessed by first perturbing their impact on the social network connectivity 
and second by using only those members, where the simulated evolution best fits 
the observed evolution.

The forecast of the COVID-19 epidemic for Slovenia, shown in 
Section~\ref{section:results}, is generated as follows. First, 1000 
perturbed simulations were performed. Among all forecasts, only 10\% of 
simulations which best follow the observed number of hospitalised patients on 
intensive care unit (ICU) and fatal (F) cases are used. The cumulative absolute 
discrepancy between observed values $y$ and modeled values $x$ at time $t_i$ is 
measured with a cost function
\begin{equation}
 J = \sum_{i=0}^N \left( \left| \log y_{ICU}(t_i) - \log x_{ICU}(t_i)\right| + 
\left| \log y_{F}(t_i) - \log x_{F}(t_i)\right| \right) .
\end{equation}
Logarithms are used to weigh equally the initial and later phase of 
the pandemic, as the number of infected varies by several orders of magnitude. 
The described data assimilation approach allow us to estimate both the impact 
of the intervention measures as well as the changes in the distribution of 
parameters.

\subsection{Exclusion experiments}
We perform exclusion experiments to assess the contribution of the above 
mentioned model components uncertainty to the total forecast uncertainty. For 
example, to estimate the contribution of the randomized social network to the 
total forecast uncertainty, we run an ensemble of simulations with the same 
social network, i.e. we exclude the social network perturbaton. 

The proxy for forecast uncertainty is the relative spread, i.e. the spread of 
the forecast ensemble, divided by the median value of the forecasts at each time 
instance. As the spread is approximately symmetric on the logarithmic axis 
(Fig.~\ref{fig:slovenia_fc}), we compute the relative spread as:
\begin{equation}\label{eq:forecast_spread}
 RS(t) = \frac{\log P_{75}(\vec{x}(t)) - \log 
P_{25}(\vec{x}(t)) }{\log\mathrm{median} (\vec{x}(t))} ,
\end{equation}
where $P_{75}$ and $P_{25}$ indicate 75th and 25th percentiles of population 
$\vec{x}$ at time $t$.

\section{Results}\label{section:results}

\subsection{Prediction for Slovenia issued on May 5, 
2020}\label{subsection:prediction}

\begin{figure}
 \centering
\includegraphics[width=\textwidth]
{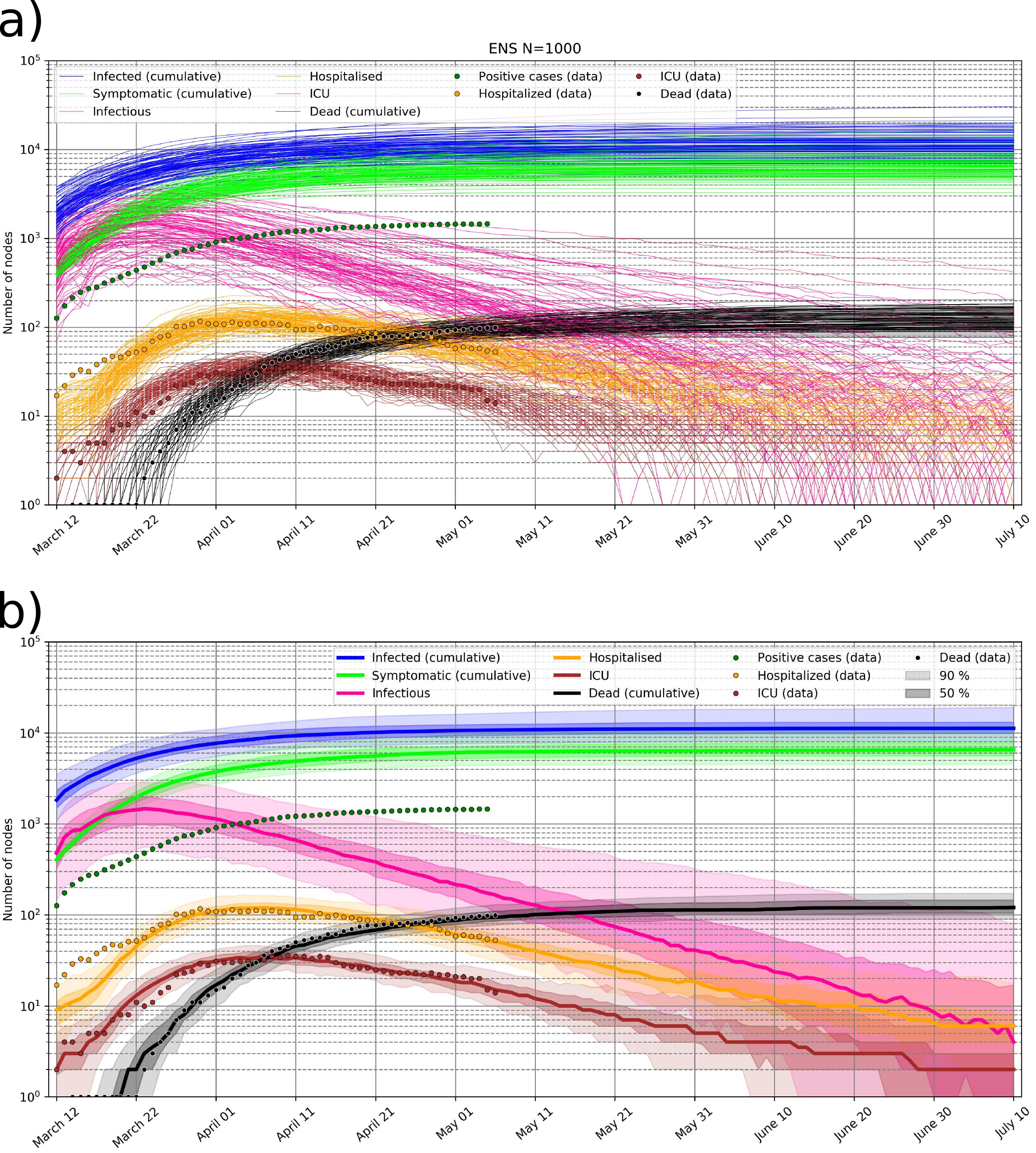}
\caption{Forecast of COVID-19 pandemic in Slovenia issued on May 5, 2020 and 
comparison with real data. a) 100 ensemble members which best fit the observed
data (dots) are shown. b) Probabilistic forecast: median value, interquartile 
range (50\%; 25th-75th percentile) and 90\% range are shown.}
\label{fig:slovenia_fc}
\end{figure}

Every day, new data is used to correct the COVID-19 forecast. 
Fig.~\ref{fig:slovenia_fc} shows an example of the ensemble prediction issued 
on May 5, 2020, simulated from the initial condition on March 12, 2020. The 
forecast is issued in the already declining stage of the epidemic and assumes 
ongoing intervention measures. Fig.~\ref{fig:slovenia_fc}a shows 100 members out 
of 1000, whose evolution least deviates from the observed data. 
Fig.~\ref{fig:slovenia_fc}b shows the associated probabilistic forecast. The 
infectious population has the largest uncertainty relative to its value, however 
the number of infectious is 
not constrained by any measurements. Thus, its relative uncertainty roughly 
reflects the uncertainty in the hospitalisation, ICU and IFR ratios. The total 
number of infected to date approaches 11000 people (90\% CI 7000-17000), in 
line with the current
estimate of the under-reporting of symptomatic cases (only 17\% of cases 
reported) by \citet{Russell2020a} and the recently estimated asymptomatic ratio 
in Italian town of Vo \citep{Lavezzo2020}.

In April 2020, a National COVID-19 prevalence survey has been 
completed \citep{slosurvey2020}, which reported 1 actively infected out of 1367 
tested (0.073\% prevalence) and 41 positive for coronavirus antibodies out of 
1318 tested (3.1\% prevalence, 95\% CI 2.2-4\%). However, the survey adds very 
little extra information to better constrain the forecast. First, the number of 
actively infected is associated with large confidence interval, and second, the 
antibody tests have significant false-positive rate and varying sensitivity 
\citep{Lassauniere2020}.

\begin{figure}
 \centering
\includegraphics[width=0.6\textwidth]
{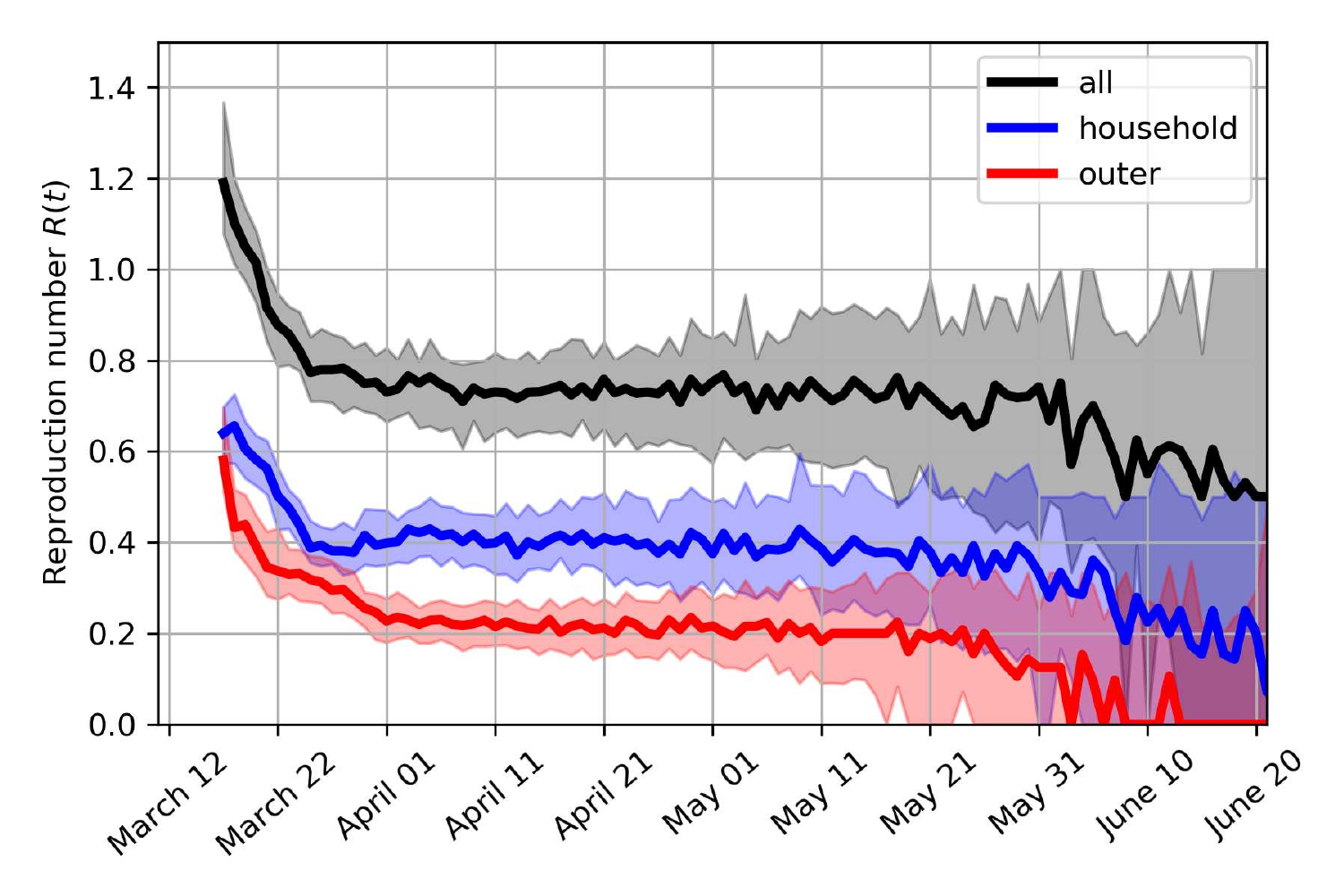}
\caption{Evolution of the estimated time-varying reproduction number $R$, 
decomposed into reproduction number associated with household transmission and 
transmission outside households. The shaded regions indicate the interquartile 
ranges.}
\label{fig:reproduction_number}
\end{figure}

In the social network model, the current reproduction number $R$ can be 
directly measured. For each infectious node, we count the number of nodes it 
infects. Then we assign the counts to the time instance corresponding to the end 
of the infectious period. Fig.~\ref{fig:reproduction_number} shows the 
reproduction number falling below 1 on March 20, 2020, which marks the 
transition into 
decaying stage of the epidemic. Current estimate of $R$ is at around 0.75, in 
line with the recent estimate for Slovenia of \citet{Manevski2020}. 
Fig.~\ref{fig:reproduction_number} also shows that the infection is currently 
much more likely to transmit within households than outside households. If the 
current intervention measures continue, the reproduction number would 
start to rapidly decline at the end of May without any extra intervention 
measures, which indicates effective virus containment when the virus would be 
transmitted only within some of the household clusters.

%

\begin{figure}
 \centering
\includegraphics[width=1\textwidth]
{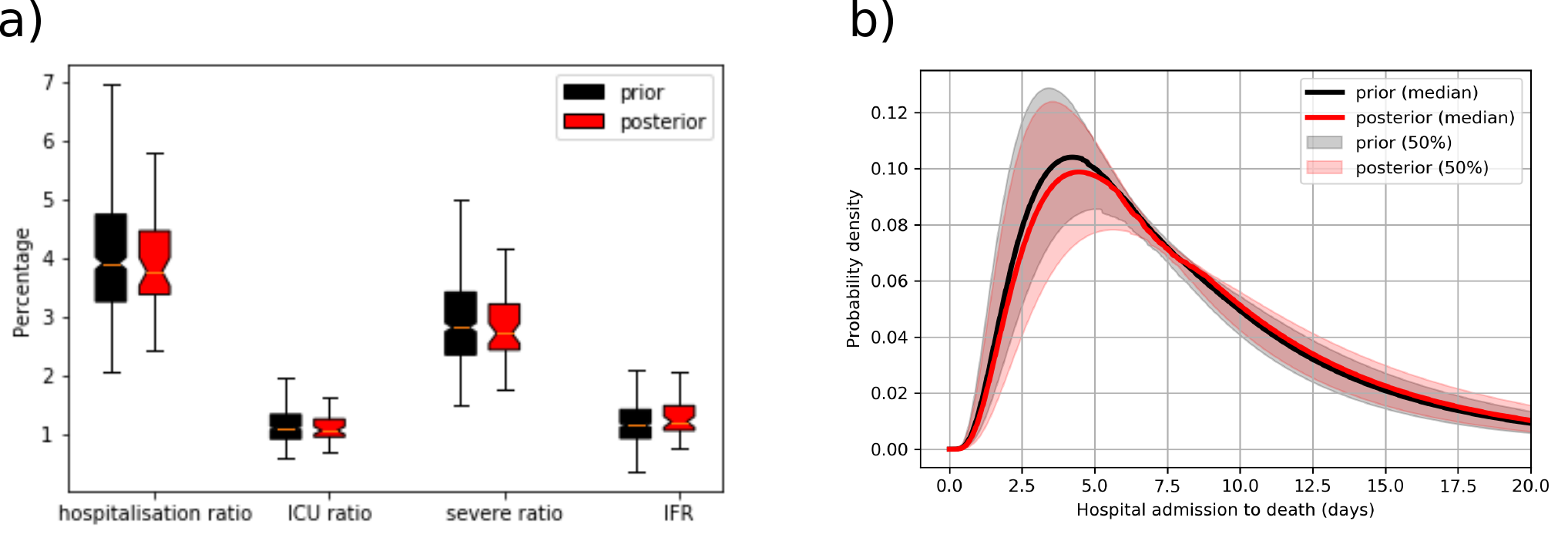}
\caption{a) Prior and posterior distributions of hospitalisation ratio, 
intensive care unit (ICU) ratio, ratio of severe symptoms (requiring 
hospitalisation), and infection fatality ratio (IFR). b) The probability 
distribution of the duration of hospital admission to death.}
\label{fig:posterior}
\end{figure}

The members of the ensemble, which minimize the cost function, can also be used 
to inverse estimate the posterior distribution of clinical parameters, 
such as hospitalisation ratio, ICU ratio, ratio of severe infection, and IFR, as 
well as disease progress parameters such as the probability distribution of the 
time-span of hospital admission to death. For 
example, according to Fig.~\ref{fig:posterior}a, the true hospitalisation 
rate is slightly smaller than the first guess, while the infection fatality 
rate is 0.1\% higher in the posterior analysis. As 
another example, Fig.~\ref{fig:posterior}b shows that the posterior estimate 
of the mean hospital admission to death duration is 7.5 days, half a day longer 
than the first guess estimate. This inverse technique was also used to estimate 
the impact of intervention measures on the social network connectivity. 
However, at the time, virus transmission parameters and some disease progress 
parameters (e.g. IFR) could not be constrained due to the lack of reliable data 
on the infectious population and total infected population.

\subsection{Forecast uncertainty decomposition}
Using the exclusion experiments, we evaluated the contribution of different 
epidemic model components to the total forecast uncertainty of the total 
number of infected and infectious population. For instance, the ensemble 
experiment where the social network and the initial condition are 
fixed (not perturbed) is termed NONET, the experiment without virus 
transmission dynamics perturbation is called NOTRANS, while the experiment 
without disease progression model perturbation is named NODIS.

\begin{figure}
 \centering
\includegraphics[width=1\textwidth]
{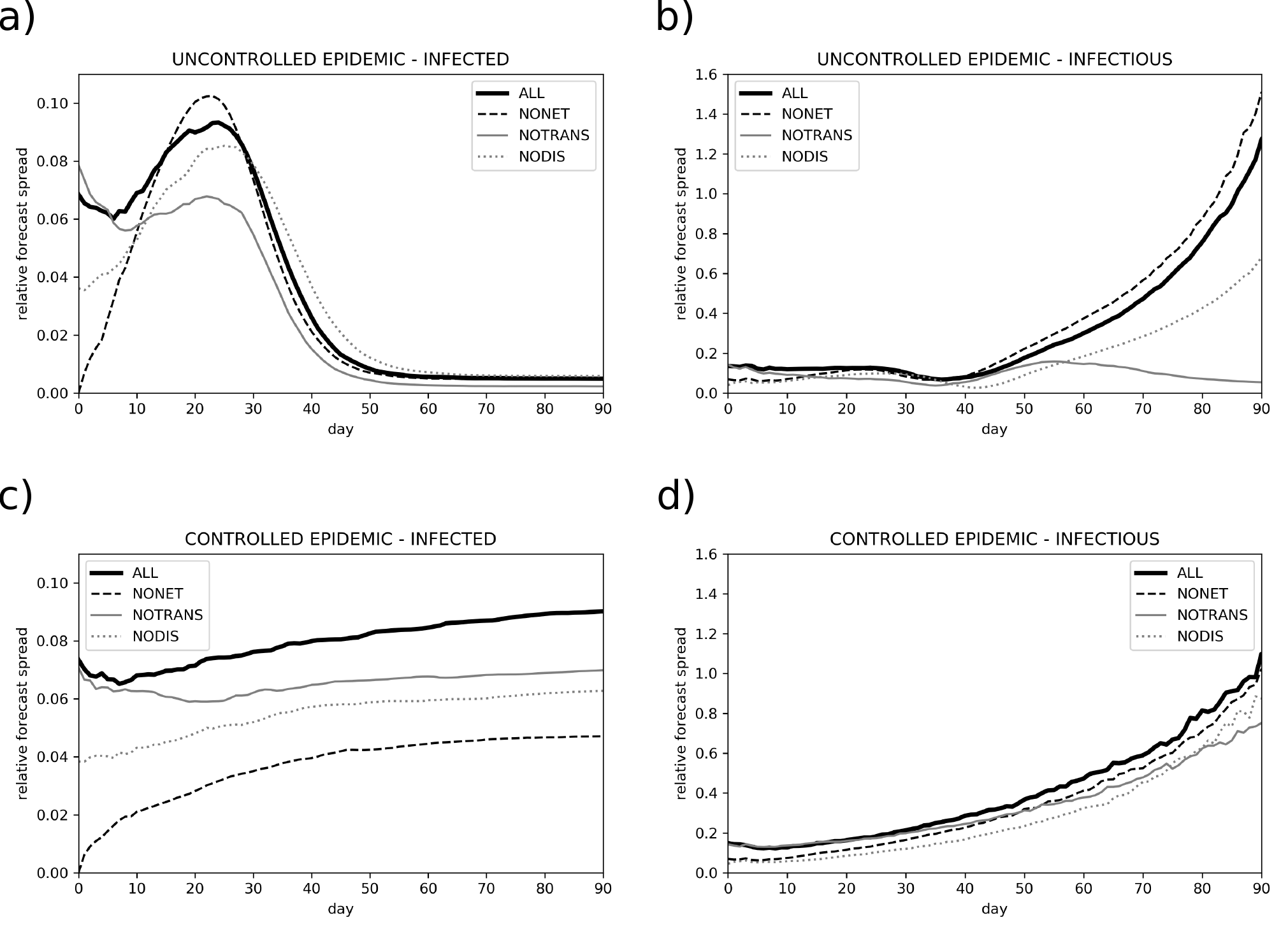}
\caption{Relative forecast spread, measured by Eq.~\ref{eq:forecast_spread}, 
for uncontrolled epidemic (a,b) and controlled epidemic with low number 
of infected (c,d), as shown in Fig.~\ref{fig:slovenia_fc}. The basic 
experiment with all parameters perturbed is termed ALL. NONET stands for 
no social network and initial condition perturbation, NOTRANS stands for 
no transmission dynamics parameters perturbations, while NODIS means 
no disease progress model parameters perturbations.}
\label{fig:uncertainty}
\end{figure}

We perform exclusion experiments for two different cases: uncontrolled epidemic 
and controlled epidemic with intervention measures and low infected population. 
The results are shown in Fig.~\ref{fig:uncertainty}. We observe, that in the 
uncontrolled epidemic, the forecast uncertainty is most reduced when the 
transmission dynamic parameters are not perturbed 
(experiment NOTRANS in Fig.~\ref{fig:uncertainty}a,b). This also reduces the 
uncertainty in the epidemic peak and later stages of the epidemic. Fixing 
disease progression parameters (such as ratio of asymptomatic infections and 
duration of infectiousness) also significantly reduces uncertainty (experiment 
NODIS). Fixing initial condition and social network structure reduces the 
uncertainty only in the initial stage of the epidemic (until around day 10), 
when the number of infected individuals is small (experiment NONET) and 
homogeneous mixing is an 
invalid assumption. In the later stage, the uncertainty becomes similar to the 
basic experiment with all parameters perturbed (experiment ALL). These 
experiments indicate that the largest contributor to the forecast uncertainty 
in the uncontrolled epidemic is virus transmission dynamics.

In the controlled epidemic with low number of infected, though, fixing the 
social network and initial condition (NONET, 
Fig.~\ref{fig:uncertainty}c) reduces the forecast uncertainty the most 
(the impact is amplified again in the initial stage), followed by fixing the 
disease progression parameters, with the impact amplified again in the early 
part of the simulation. This suggests that the structure of the network 
and the initial distribution of infected nodes drastically affects the 
evolution due to heterogeneous mixing and randomized irregular social network. 
The result suggests that the epidemic forecast can be improved (i.e. its 
uncertainty decreased) the most by constructing a more realistic model of our 
social network.

\section{Dicussion, conclusions and further outlook}

In this study, we have developed a virus transmission model on the simplified 
social network of Slovenia with 2 million nodes organised into home/care 
center clusters. A detailed disease progression model was coupled with the 
virus transmission model. The model probabilistic prediction is regularly 
updated on Sledilnik webpage \citep{sledilnik2020} and is occasionally 
communicated to the Expert Group that provides support to the Government of the 
Republic of Slovenia for the containment and control of the COVID-19.

We have developed a data assimilation procedure, which minimizes the cost 
function measuring the deviation from observed ICU, hospitalisation and fatality
numbers. The procedure constrains the forecast trajectories closer to the 
observed values. It also constrains the model parameters. Our approach mimics 
the established variational data assimilation approach in Numerical Weather 
Prediction (NWP) \citep{Kalnay2003,Lahoz2014}. Another example of data 
assimilation utilisation in epidemiology is by \citet{Shaman2012}, who used an 
Ensemble Adjustment Kalman Filter \citep{Anderson2001}.

An indispensable part of the prediction is its uncertainty. In this study, we 
evaluated the contribution of the virus transmission uncertainty (e.g. 
reproduction number and its derivatives), network and initial condition 
uncertainty and uncertainty of the disease progress model to the total 
uncertainty of the epidemic forecast. We found that in the uncontrolled 
epidemic, the intrinsic uncertainty mostly originates from the uncertainty of 
the virus transmission. In the controlled epidemic with low infected 
population, the randomness of the social network becomes the major source of 
forecast uncertainty. We also show, that the uncertainty of the forecast and 
the associated risk is extremely asymmetric (roughly symmetric on a 
logarithmic axis) with long exponential tails, reaching a similar conclusion to 
the recent study of \citet{Petropoulos2020}.

There are some limitations of our model which reduce its predictive ability 
and its usefulness to simulate the impact of intervention measures in advance. 
The social network model is too simplified, and its average clustering too low. 
For example regional, work/education clustering based on work/education 
mobility data is not included in the present social network. The nodes do not 
have attributions such as age, sex or employment status and the social mixing 
data \citep{Mossong2008,Meyer2017,Prem2017,Davies2020} is not accounted for 
yet. Given the high attack rate within households, the social mixing within 
households is of special importance, thus it is also vital to include the 
age-distribution of the residents of different household sizes. A more 
sophisticated treatment of the secondary attack rate is also needed, for example 
the infectiousness could be modeled as a function of time \citep{He2020}. 
Further work should alleviate some of the mentioned limitations to allow more 
robust simulation of the intervention measures.

The ongoing COVID-19 epidemic has revealed a major gap in our ability to 
forecast the evolution of the epidemic. No operational center for infectious 
disease prediction, similar to those employed for the weather 
predictions (e.g. European Centre for Medium-Range Forecasts or National Center 
for Environmental Prediction), exists, despite the gigantic societal, 
economical and health impact of the ongoing epidemic. While the epidemic 
dynamics is governed by the human social behaviour and its modeling arguably 
messier than weather forecasting \citep{Moran2016}, a coordinated modeling 
effort which borrows the established methods used for Numerical Weather 
Prediction (NWP) would likely improve our prediction \citep{Shaman2012}. 
Accurate models of the real-world social networks are needed to realistically 
simulate the virus transmission dynamics. Similarly to NWP models 
\citep{Bauer2015}, the real-time clinical patient data, 
mobility data \citep{Gonzalez2008} and connectivity data (obtained by e.g. 
postprocessing the bluetooth-generated anonymous contact data 
\citep{Dandekar2020}), should be rapidly assimilated into the virus spread 
prognostic model \citep[e.g.][]{Cao2020} to evaluate the changes in contact 
patterns \citep{Zhang2020}. This would allow 1) to estimate the critical virus 
spread parameters and their uncertainty, 2) to forecast the evolution of the 
epidemic more accurately and based on that forecasts, 3) to implement optimal 
worldwide-concerted measures to minimize the virus spread. We should be ready 
for the next big pandemic!

\section*{Acknowledgments}
The authors are grateful to Aleks Jakulin, Miha Kadunc (Sinergise), Luka Renko 
(founder of the volunteer-led Sledilnik.org project) for providing the timely 
COVID-19 pandemics data for Slovenia. Žiga Zaplotnik would like to thank prof. 
Alojz Kodre and asis. prof. Simon Čopar (both University of Ljubljana, Faculty 
of Mathematics and Physics, UL-FMF) for introducing him the node-based analysis 
of the information spread. We thank fellow physicists Nejc Davidovič and Jan 
Bohinec (Gen-I) for fruitful discusions as well as asis. prof. Samo Drobne 
(University of Ljubljana, Faculty of civil and geodetic engineering) for 
providing the mobility data for Slovenia. Arthur Breitman (Tezos) is thanked for 
an interesting question on the properties of the virus transmission dynamics 
related to superspreaders, which initiated further research. Special thanks go 
to prof. Roman Jerala (National Institute of Chemistry) and prof. Tomaž Zwitter 
(UL-FMF) for discussions of the model, and for reading and commenting the 
early manuscript draft as well as for providing useful literature. Finally, 
Žiga Zaplotnik would like to thank his postdoc supervisor prof. dr. Nedjeljka 
Žagar (University of Hamburg) for always supporting work beneficial to society.

\bibliography{library}


\end{document}